# Liquidity on Web Dynamic Network

*Massimiliano Dal Mas*
*me @ maxdalmas.com*

**ABSTRACT**
Nowadays, the exponentially growing of the Web renders the problem of correlation among different topics of paramount importance. The proposed model can be used to study the evolution of network depicted by different topics on the web correlated by a dynamic "fluid" of tags among them. The fluid-dynamic model depicted is completely evolutive, thus it is able to describe the dynamic situation of a network at every instant of time. This overcomes the difficulties encountered by many static models. The theory permits the development of efficient numerical schemes also for very large networks. This is possible since dynamic flow at junctions is modeled in a simple and computationally convenient way (resorting to a linear programming problem). The obtained model consists of a single conservation law and is on one side simple enough to permit a complete understanding, on the other side reach enough to detect the evolution of the dynamic network.

## INTRODUCTION

There are many motivations for the study of tags correlation in a dynamic environment. However, modeling flow of tags is a non-trivial task, and tags interaction on different topics in particular are difficult to reproduce.

Dynamic correlation on tags can be treated by different models. [1-7] We focus on macroscopic fluid-dynamic models. [8]

Our reference mathematical model is based on conservation laws and represents a network as a finite set of paths that are modeled by intervals and connected to each other by junctions. It is schematized as an oriented graph, where topics are modeled as arcs and junctions as nodes.

One of the strengths of this method is the low level of information needed to detect the load on the whole network.

From a numerical point of view, the algorithm is based on approximation methods and kinetic schemes with suitable boundary conditions at junctions.

A description of the mathematical model proposed is followed by the experiment done. Finally, some conclusions and future works are introduced.

---

## THE BASIC IDEA

On the web different topics are correlated to different tags by the users. These correlations can change and tags can be correlated to different topics forming a Dynamic Network that can change according to the flow of tags form one topics to another.

On this work the "liquidity" on a Dynamic Network is depicted considering tags as flow from different source S as topics.

## NETWORK FLOW

Here we refer to the book of Haberman [2] for the description of the mathematical model of network flows.

Modelling the dynamics of social networks is therefore of crucial importance, but it is also extremely difficult, due to the temporal dependence. [9-12]

On this work we focus our attention on the network situations resulting from the complex interaction of many tags, instead of studying the behaviour of individual tags.

The study of network problems proposes to answer to several questions. The principal aim is to discover flow phenomena of the tags in order to determine what topic is of more interest. [2, 11, 12]

We will only formulate deterministic mathematical models, but it is also possible to develop statistical theories. The treatment of these problems is based on the fundamental network variables of the tags: *velocity*, *density* and *flow*. The nonlinear partial differential equation (1) represents the relationships between tag appearance velocity and network density.

$$(1) \quad \partial_t \rho + \partial_x f(\rho) = 0$$

## Velocity Field

With $N$ tags there are different velocities, $v_i(t)$, $i = 1, \ldots, N$, each depending on time. If the number of tags $N$ is large, it becomes difficult to keep track of each tag. So, instead of measuring the velocity of each individual tag, we associate to each point in space at each time a velocity field, $v(x,t)$. This would be the velocity measured by an observer at time t.

## Network Flow and Density

In addition to the tags velocities we could measure the number of tags that appears in a given length of time. The average number of tags appearing per time unit (for example one minute) is called the network flow $q = q(x,t)$. A systematic procedure could be used to take into account tags completely in a given region defined by a topic at a fixed time; estimates of fractional tags could be used or a tag could be counted only if its clearly related to a topic. These measurements give the density of tags, $\rho$, that represents the number of tags per topic – respect all the tags owned by an ontology.

## Flow equals density times velocity

There is a close relationship between the three fundamental network variables: velocity, density and flow. It is quite realistic to think to the flux $q$ - the number of tags per time unit - as a function of the only density $\rho$. More precisely the flux will be expressed as:

*network flow = (network density) × (mean velocity of tag appearance)*

That can be written as in (2)

$$(2) \quad q(x,t) = \rho(x,t) v(x,t)$$

As the density increases (meaning there are more and more tags per subject), the velocity of appearance of tags diminishes. Thus we make the hypothesis that the velocity of tags at any point is a regular strictly decreasing function of the density:

$$v = v(\rho)$$

## An example of Liquid Network Dynamic: Circle of Tags

A "circle" is a type of looping junction in which flow travels in one direction around a central island and priority is given to the circulating flow.

The fundamental principle of circle is that entering flow can go out in different paths. Considering the paths as the topic *S* and the flow as the tags we have that entering flow of tags relative to an ingoing topic (eg. *S1*) could be associated to a different outgoing topic (eg. *S4*). So analyzing the flow of the tags on the circle it's possible determine the correlation between the topics (eg. correlation between the topic *S1* and *S2*).

In this section we fix a simple network representing a flow circle of tags and assume that there is low flow, in the sense that the number of tags reaching the circle is less then the capacity of the circle itself.
For our example we consider four topics, named *S1*, *S2*, *S3*, *S4*, the first two incoming in the circle and the other two outgoing. Moreover there are four correlated topics *S1C*, *S2C*, *S3C*, *S4C* that form the circle as in Figure 1

**Figure 1**  Tags flow circle

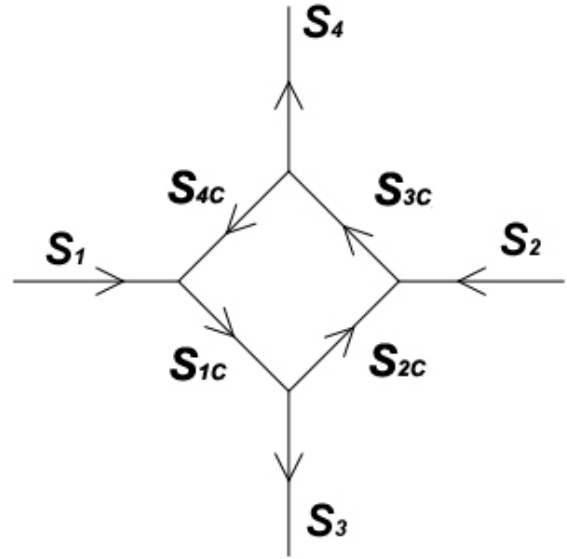

Topics *S* are parameterized by *[ai , bi]*, *i* = 1, . . . , 4, and *[aiC, biC]* with *i = 1, . . . , 4*. It is natural to assign a flow distribution matrix to describe how flow coming from topics *S1*, *S2* choose to be forgiven for topics *S3* and *S4*. Indeed the topics of the circle are just intermediate towards the final topics. Thus we assume to have two fixed parameters: $\alpha, \beta \in ]0.1[$
So that:
- If *N* tags reach the circle from topic *S1*, then *αN* go to topic *S3* and *(1 − α)N* go to topic *S4*;
- If *N* tags reach the circle from topic *S2*, then *β N* go to topic *S4* and *(1 − β)N* go to topic *S3*.

To fix boundary conditions we consider a static situation with constant densities from the topics *S1* and *S2* respectively:

(3) $\quad \rho_1(t, x_1) \equiv \overline{\rho}_1 \quad , \quad \rho_2(t, x_2) \equiv \overline{\rho}_2$

If the outgoing topics *S3* and *S4* can absorb all incoming flow form tags as in (4) then we should reach the situation of Figure 2, where the constant fluxes on each topic are written.

$$(4) \quad f(\overline{\rho}_1) + f(\overline{\rho}_1) \leq f(\sigma)$$

For this to happen we must have that the coefficients for the crossing (*S1C*, *S3*, *S2C*) are depicted in (5) and (6).

$$(5) \quad \alpha_{1R,3} = \frac{\alpha f(\overline{p}_1) + (1-\beta)f(\overline{p}_2)}{f(\overline{p}_1) + (1-\beta)f(\overline{p}_2)}$$

$$\alpha_{1R,2R} = \frac{(1-\alpha)f(\overline{p}_1)}{f(\overline{p}_1) + (1-\beta)f(\overline{p}_2)}$$

$$(6) \quad \alpha_{1R,3} = \frac{(1-\alpha)f(\overline{p}_1) + \beta f(\overline{p}_2)}{(1-\alpha)f(\overline{p}_1) + f(\overline{p}_2)}$$

$$\alpha_{3R,4R} = \frac{(1-\beta)f(\overline{p}_2)}{(1-\alpha)f(\overline{p}_1) + f(\overline{p}_2)}$$

If the network is initially empty, the boundary data are given by (1) and (2) holds true, then firstly the tags from topic *S1* and *S2* reach topic *S3* and *S4* respectively and the coefficients should be simply set as:

$$(7) \quad \alpha_{1R,3} = \alpha, \quad \alpha_{1R,2R} = (1-\alpha)$$
$$\alpha_{3R,4} = \beta, \quad \alpha_{3R,4rR} = (1-\beta)$$

However, then also tags from topic *S2* reach topic *S3* (and tags from topic *S1* reach topic *S3*) then we should modify in time the coefficients and finally set as in (5) and (6). With this choice, there exists $T > 0$ such that the solution is given by the fluxes indicated in Figure 2 for every $t \geq T$. Thus we see that the model is working properly at not too heavy flow level, i.e. for which (4) holds true. However, it is necessary to let the coefficients α vary on time, more precisely.

**Figure 2** Equilibrium for flow circle

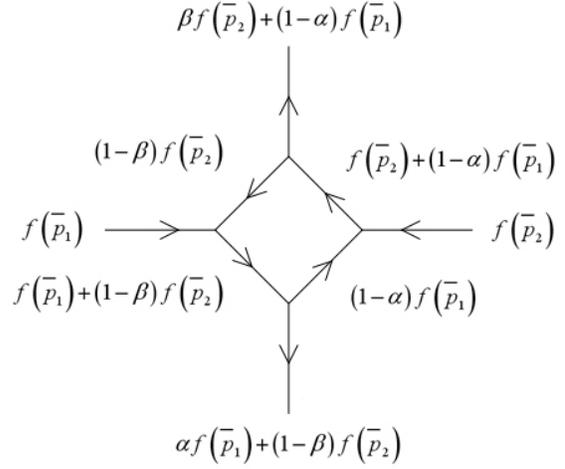

Consider the circle network and assume (3), (4). There exists time dependent coefficients α : [0, +∞) → [0, 1], with (7) holding at time 0 and (5), (6) for large enough times, and $T > 0$ such that the solution $\rho(t)$ is constantly equal to that of Figure 2 for every $t \geq T$.

To prove the assumptions made we construct solution by wave front tracking. Tags from topics *S1* and *S2* reach outgoing topics *S3* and *S4* via rarefaction fans. At each time a rarefaction shock reach either the junction (*S1C*, *S2C*, *S4*) or the junction (*S3C*, *S4C*, *S3*) we adjust the corresponding coefficients in such a way that tags reach the correct topic.

**EXPERIMENT**

We have built a simulation model reproducing the behavior of tags within a network of different topics. The simulator prototype can describe the evolution of flow of tags over time taking interactions at junctions into consideration. We have investigated the numerical validity of the approximation algorithm. It has proven to be very fast, providing solutions on networks with a few thousands arcs in less than a second of CPU time on a PC.

We then tested the model's effectiveness using subset of twitter with 20.000 tweets considered as topics S (arcs) and the flow represented by the #hashtags that correlate different topics. [13, 14]

The fast numerical algorithms described above render the relative optimization problem treatable. This procedure is based on the comparison between measured data and the simulated solutions produced by the algorithm, and gives an average percentage error of around 16%.

## CONCLUSION

In the future, we intend to test the calibration procedure on a more completed network, and to construct a real-time interface for the automatic representation and visualization of density curves.

## References

[1] M. Dal Mas, "Modelling web user synergism for the similarity measurement on the ontology matching: reasoning on web user felling for uncertain evolving systems". Proceedings of the 3rd International Conference on Web Intelligence, Mining and Semantics, Article No. 17, ACM New York, NY, USA (http://dl.acm.org/citation.cfm?doid=2479787.2479799 ) ISBN: 978-1-4503-1850-1 DOI:10.1145/2479787.2479799

[2] J. Euzenat, P. Shvaiko. Ontology Matching Springer-Verlag, Berlin (Germany), 2010 (http://book.ontologymatching.org)

[3] M. Dal Mas, "Folksodriven Structure Network". Ontology Matching Workshop (OM-2011) collocated with the 10th International Semantic Web Conference (ISWC-2011), CEUR WS vol. 814 (http://ceur-ws.org/Vol-814), 2011

[4] M. Dal Mas, "Elastic Adaptive Ontology Matching on Evolving Folksonomy Driven Environment" in Proceedings of IEEE Conference on Evolving and Adaptive Intelligent System (EAIS 2012), Madrid, Spain, 35-40, IEEE, (http://ieeexplore.ieee.org/xpl/articleDetails.jsp?arnumber=6232801) DOI: 10.1109/ EAIS.2012.6232801

[5] M. Dal Mas, "Intelligent Interface Architectures for Folksonomy Driven Structure Network" in Proceedings of the 5th International Workshop on Intelligent Interfaces for Human-Computer Interaction (IIHCI-2012), Palermo, Italy, 519– 525, IEEE, 2012 (http://ieeexplore.ieee.org/xpl/articleDetails.jsp?arnumber=6245653) DOI: 10.1109/CISIS.2012.158

[6] M. Dal Mas, "Elasticity on Ontology Matching of Folksodriven Structure Network". Accepted for the 4th Asian Conference on Intelligent Information and Database Systems (ACIIDS 2012) - Kaohsiung Taiwan R.O.C., 2012, CORR – Arxiv (http://arxiv.org/abs/1201.3900)

[7] M. Dal Mas, "Elastic Adaptive Dynamics Methodology on Ontology Matching on Evolving Folksonomy Driven Environment". Journal

Evolving Systems 5(1): 33-48 (2014) (http://link.springer.com/article/10.1007%2Fs12530-013-9086-5)

[8] R. Haberman, "Mathematical models." Prentice-Hall, Inc. New Jersey, 1977, 255-394.

[9] I. Vragovic, E. Louis, A. Diaz-Guilera, "Efficiency of information transfer in regular and complex networks.", Phys Rev E 71:026122, 2005

[10] U. Brandes, T. Erlebach, "Network analysis: methodological foundations". Lecture notes in computer science, vol 3418. Springer, New York, 2005

[11] S.P. Borgatti, A. Mehra, D.J. Brass, G. Labianca, "Network analysis in the social sciences". Science 2009 323(5916):892–895

[12] G. Bretti, R. Natalini, B. Piccoli, A Fluid-Dynamic "Traffic Model on Road Networks", Archives of Computational Methods in Engineering 14 (2007), 139-172;

[13] E. Aramaki, S. Maskawa, and M. Morita. "Twitter catches the flu: Detecting influenza epidemics using twitter". In Proceedings of the Conference on Empirical Methods in Natural Language Processing, pages 1568–1576. Association for Computational Linguistics, 2011.

[14] A. Hughes and L. Palen. "Twitter adoption and use in mass convergence and emergency events". International Journal of Emergency Management, 6(3):248–260, 2009